\documentclass[aps,prb,twocolumn,showpacs,amsmath,amssymb]{revtex4-1}

\usepackage{graphicx}
\usepackage{subfigure} 
\usepackage{dcolumn}
\usepackage{bm}

\newcommand{\bk}{{\bm k}}

\begin{document}

\title{Surface polar optical phonon interaction induced many-body
  effects and hot-electron relaxation in graphene}
\author{E.~H.~Hwang$^{1,2}$ and S.~Das Sarma$^{1}$}
\address{$^1$Condensed Matter Theory Center, 
Department of Physics, University of Maryland, College Park,
Maryland  20742-4111} 
\address{$^2$SKKU Advanced Institute of Nanotechnology,
Sungkyunkwan University, Suwon 440-746, Korea}
\date{\today}

\begin{abstract}

We theoretically study various aspects of the electron-surface optical
phonon interaction effects in graphene on
a substrate made of polar materials. 
We calculate the electron self-energy in the presence of the surface phonon-mediated
electron-electron interaction
focusing on how the linear chiral graphene dispersion is renormalized
by the surface phonons. 
The electron self-energy as well as the quasiparticle
spectral function in graphene are calculated, taking into account
electron-polar optical phonon interaction 
by using a many body perturbative formalism.
The scattering rate of free electrons due to polar interaction with
surface optical phonons in a dielectric substrate is calculated as a function of the
electron energy, temperatures, and carrier density. 
Effects of screening on the self-energy and scattering rate are discussed. 
Our theory provides a comprehensive quantitative (and qualitative)
picture for surface phonon interaction induced many-body effects and
hot electron relaxation in Dirac materials.


\end{abstract}

\maketitle

\section{introduction}

Lattice vibrations are inevitable intrinsic sources of carrier 
scattering in semiconductors and often dominate transport near room
temperature \cite{ando1982,dassarma2011}. 
In general, scattering by acoustic phonons is the main contribution to
the electronic resistivity at
low temperatures, and the longitudinal-optical (LO) phonons in polar
materials limit the room temperature mobility 
through their coupling with carriers via the long-range
Fr\"{o}hlich interaction \cite{mahan}.
Since graphene is made of carbons it is a non-polar material and
the lack of strong long-range polar optical phonon scattering
leads to very high intrinsic (i.e., phonon-limited) room-temperature
graphene mobility \cite{morozov2008,chen2008b,zou2010,hwang2008,
  fratini2008,min2011,efetov2010} 
and indeed the room-temperature graphene mobility is typically very
high in disorder-free suspended graphene samples \cite{du2008,dassarma2013}. 
The weak deformation potential scattering from the thermal lattice
acoustic phonons mainly limits the intrinsic mobility of graphene at room temperature
and the acoustic phonon scattering gives a quantitatively small contribution
even at room temperature due to the high Fermi temperature 
at high carrier density (and the small deformation potential
coupling) of graphene \cite{hwang2008,min2011,efetov2010}.

However, most current available graphene samples 
for fundamental studies and certainly for technological applications
use
a polar substrate such as SiO$_2$, SiC, or HfO$_2$.  
Such polar substrates allow for the existence of polar optical phonons
localized near the graphene-substrate interface which could be an
important scattering source for graphene carriers through the
long-range Fr\"{o}hlich coupling.
The general eigenvalue equation for a surface phonon mode has been derived 
at a planar interface between two polar dielectrics
\cite{fuchs1965,economou1969,wang1972}. 
Since the surface optical (SO) phonon is a
well-defined surface property at the interface of the polar
semiconductors \cite{wang1972,hess1979} it is
possible that carriers in graphene layer couple to the SO phonon of
the underlying substrate lattice via the long-range polar Fr\"{o}hlich
coupling. 
Recently, the possible role of surface optical phonon on high temperature
graphene mobility has been pointed out
\cite{chen2008b,zou2010,fratini2008}. Due to the polar 
nature of the gate dielectrics used in 
graphene field effect transistors,
the carriers in the conducting channel electrically couple to the
long-range surface phonons created at the dielectric interface.
In general, the surface optical phonon 
contributes little to the carrier resistivity at low
temperatures due to its high energy ($\omega_{SOP} \agt 50
meV$). However, the contributions of other scattering processes to the
graphene resistivity  
are small at room temperature and therefore the SO phonon may become 
the dominant scattering mechanism in graphene on a polar dielectric
substrate at high temperatures.  It is therefore possible, perhaps
even likely, that in the context of graphene technological
applications, SO phonon effects are particularly important and must be taken
into account. 
In addition, in a recent experiment \cite{liu2010,lu2009,koch2010,yan2012}, the
strongly coupled plasmon-phonon mode dispersion has been measured by the
angle-resolved reflection electron-energy-loss spectroscopy and it is
found that the electron-phonon coupling give rise to the many
interesting many-body effects in graphene on a polar dielectric
substrate \cite{hwang2010}.
This again points to the possible importance of carrier-SO phonon coupling
in determining the electronic properties of graphene on a substrate.

Since the electron-optical phonon interaction leads to many-body
renormalization of the single-particle free carrier properties
\cite{tse2007,park2007} while at the same affecting the transport
properties \cite{fratini2008,chen2008b}  a good understanding of
electron-SO phonon coupling is thus important in developing quantitative
theories for many different experimental studies in graphene.  
In this paper we calculate the electron self-energy and spectral
function of graphene on a polar substrate such as
SiO$_2$, SiC, or HfO$_2$.  
Our goal in this paper is to provide a fairly detailed theoretical
picture for the electron-SO phonon interaction induced polar
scattering of free carriers. Many electronic
properties of graphene on polar substrates are expected to be
modified by electron-SO phonon scattering via the Fl\"{o}hlich
interaction. On a microscopic level this interaction renormalize the
Fermi velocity of graphene.
A thorough quantitative understanding of the electron-SO phonon
interaction induced many-body effects in high mobility graphene is
very important with respect 
to both the device operation and the fundamental physics of graphene
on substrates. This surface polaronic (i.e. interaction with SO
phonons) effect is an important factor in determining
the high temperature mobility and the hot electron properties of graphene
on substrates.

In this paper, we focus on the effect of electron-SO phonon
interaction effects on the real, Re[$\Sigma$], and the imaginary part,
${\rm Im}[\Sigma]$, of electron self-energy.
From ${\rm Im}[\Sigma]$, we can extract the quasiparticle scattering
rate (scattering lifetime),
which gives information relevant to possible applications,  
such as the hot electron relaxation rate and the
energy dissipation rate of injected carriers in a
graphene-based device.  
We calculate the scattering rate as a functions of wave vector, energy, density and
temperature and discuss the role of screening effects. Since the electron-
surface polar optical phonon interaction has a long range nature the
consideration of screening effects is crucial in order to understand
quantitative physical properties of graphene.
We find that the scattering rates with screened interaction is reduced
by a factor of five, compared with the results with unscreened
interaction. Our consideration of screening (i.e., the static random
phase approximation (RPA) screening) may overestimate the
screening  effect on scattering rate at low temperatures, but it
provides an upper limit of the screening effects. 
In addition to calculating the properties of Im[$\Sigma$] due to SO
phonon interaction, we also calculate the Re[$\Sigma$], which gives us the
corresponding graphene velocity renormalization due to the surface
polaronic effect.  We also provide results for the graphene spectral
function due to electron-SO phonon scattering, which can be
directly observed in scatting tunneling spectroscopy (STS) and
angle-resolved photo-electron spectroscopy (ARPES).  Our theory and
results are directly relevant to the electronic properties of doped or
gated graphene on polar substrates where SO polaron effects may
very well be significant. 


The paper is organized as follows. In Sec. II the generalized theory
is presented to calculate the self-energy of electron in the presence
of electron-SO phonon interaction.
Sec. III presents the results of self-energy and spectral function.
We also show the scattering rate as a function of energy, momentum, density,
and temperature. We summarize in Sec. IV with a discussion.

\section{theory}

The Hamiltonian of graphene is well-approximated by a 2D
Dirac equation for massless particles, ${H}_0 = v_F \sigma \cdot {\bm k}$,
where $v_F$ is the Fermi velocity of graphene, $\sigma$ is the
pseodospin $2\times 2$ Pauli matrices, and 
${\bm k}$ is the 2D momentum relative to the Dirac points (we use $\hbar = 1$
throughout this paper).  
The two components of the spinors correspond to occupancy of the two
sublattices of the honeycomb structure in a hexagonal lattice.   
This ${H}_0$ gives a linear energy dispersion relation
$\varepsilon_{\bm k,s} = s  v_F |{\bm k}|$,  
where $s = +1$ $(-1)$ for the conduction (valence) band.
The corresponding density of states (DOS) is 
$ D(\varepsilon) = g_s g_v |\varepsilon|/(2\pi v_F^2)$, where
$g_s=2$, $g_v=2$ are the spin and valley degeneracies,
respectively.

The central quantity we calculate in this paper is the leading order
electronic self-energy correction due to electron-surface optical (SO)
phonon interaction via Fr\"{o}hlich polar coupling \cite{wang1972}.
In the presence of the long range electron-SO phonon
coupling, electrons interact among themselves through the Coulomb
interaction and through virtual-SO phonon exchange via the
Fr\"{o}hlich interaction.  The electron-SO phonon interaction is given
by
\begin{equation}
H_{e-ph}=\sum_{kq}\sum_{ss'}M_{kq}^{ss'}c_{k+qs'}^{\dagger}c_{ks} \left (
b_q + b_{-q}^{\dagger} \right ),
\end{equation}
where $c_{ks}^{\dagger}$ is the electron ($s=+1$) or hole ($s=-1$)
creation operator, $b_q^{\dagger}$ and $b_q$ are creation and
destruction operators of surface optical phonon, and the interaction matrix
element $M_s^{ss'}$ is defined by
\begin{equation}
M_{kq}^{ss'}=M(q)F_{sk+q}^{\dagger}F_{s'k},
\end{equation}
where $F_{sk}$ is the chiral spinor and given by 
\begin{equation}
F_{sk}=\frac{1}{\sqrt{2}}
\left ( \begin{array}{c}
s \\
e^{i\theta_k} \end{array}
\right )
\end{equation}
with $s=\pm1$, $\theta_k=\tan^{-1}(k_y/k_x)$ 
is the angle between $\bm{k}$, $\bm{k}'$, 
arises from the overlap of $\mathinner|\!s{\bm k}\rangle$ and
$\mathinner|\!s'{\bm k}'\rangle$. \cite{dassarma2011}
The strength of the coupling $M(q)$ is given by
\begin{equation}
[M(q)]^2 = \frac{M_0^2}{q\epsilon(q)}e^{-2qd}
\label{eq:mq}
\end{equation}
where 
\begin{equation}
M_0^2 = \pi e^2 \omega_{SO} \left [
  \frac{1}{\epsilon_{\infty}+1} - \frac{1}{\epsilon_0+1} \right ],
\end{equation}
and $\epsilon(q)$ is the dielectric (screening) function of graphene
\cite{hwang2007}, $d$ 
is the separation distance between graphene layer and  
substrate, $\omega_{SO}$ is the surface optical phonon frequency,
and $\epsilon_0$
($\epsilon_{\infty}$) is the static (high frequency) dielectric constant. 
In Eq.~(\ref{eq:mq}), we set $\epsilon(q) = 1$ for the unscreened
case. For screened interaction we use the static RPA dielectric
function appropriate for graphene, which is give by
\begin{equation} 
\epsilon(q) = 1 + v(q)\Pi(q),
\end{equation}
where $v(q)=2\pi e^2/\kappa q$ is the electron-electron Coulomb potential
with the effective background lattice
dielectric constant $\kappa = (\epsilon_0 + 1)/2$, and
$\Pi(q) = \Pi(q,T)$ is the electronic polarizability function of
graphene which depends on both wave vector and temperature. 
For isotropic media the frequency of SO
phonons $\omega_{SO}$ is related to the transverse optical (TO) bulk
phonon $\omega_{TO}$ as
${\omega_{SO}}/{\omega_{TO}}=\sqrt{({\epsilon_0+1})/({\epsilon_{\infty}
    +1})}$.\cite{fuchs1965,wang1972}
Note that the bulk longitudinal optical phonons $\omega_{LO}$ 
and $\omega_{TO}$
are connected with the dielectric constants by the
Lyddane-Sachs-Teller relation \cite{mahan}
${\omega_{LO}}/{\omega_{TO}} = \sqrt{
  {\epsilon_0}/{\epsilon_{\infty}}}$.


The SO-phonon mediated electron-electron effective interaction 
(i.e. in the leading-order screened Fr\"{o}hlich coupling) is given by
\begin{equation}
v_{ph}^{ss'}(q,\omega)=[M_{0}(q)]^2D_0(\omega)F^{ss'}_{kk'},
\label{v_phss}
\end{equation}
which depends on both wave vector and
frequency. In Eq.~(\ref{v_phss}) $D_0(\omega)$ is the unperturbed
SO-phonon propagator and given by
\begin{equation}
D_0(\omega)=\frac{2\omega_{SO}}{\omega^2-\omega_{SO}^2}
\end{equation}
and
\begin{equation}
F^{ss'}_{kk'} = \left | F_{sk}^{\dagger} F_{s'k+q} \right |^2 =
  {[1+ss'\cos(\theta_{kk'})]}/{2}.
\end{equation}

The self-energy $\Sigma$ within the leading-order 
approximation is give by \cite{mahan} 
\begin{equation}
\Sigma_s(\bm k,i\omega_n) =
-\frac{1}{\beta}\sum_{s'}\sum_{{\bm q},i\nu_n} [M^{ss'}_{kq}]^2D(q,i\omega_n)
G_{0,s'}(\bm k+\bm q,i\omega_n+i\nu_n).
\label{sigma}
\end{equation}
Here, $\beta = 1/k_BT$, $s,s'=\pm 1$ are band indices, 
$G_{0,s}({\bf k},i\omega_n)=1/(i\omega_n-\xi_{{\bf k}s})$
is the bare Green's function, $\omega_n,\nu_n$ are Matsubara fermion
frequencies, 
and $M_{kq}^{ss'}$ the interaction matrix element.
After the standard analytical continuation from $i\omega_n$ to
$\omega+i0^+$ the retarded self-energy $\Sigma^{\rm ret}$ is obtained,  
\begin{widetext}
\begin{equation}
\Sigma_s^{\rm ret}(\bm k,\omega) =  \sum_{s'}\int\frac{d\bm
  q}{(2\pi)^2} \frac{M_0^2 e^{-2qd}}{q \epsilon(q)}
F_{ss'}(k,k-q) \left [ \frac{N_0 + 
      n_F(k_{s'})}{\omega + i \eta + \omega_{SO} - \xi_{k's'}} + \frac{N_0 +1
      -n_F(k_{s'})}{\omega + i\eta -\omega_{SO}-\xi_{k's'}} \right ]
\label{eq:2}
\end{equation}
\end{widetext}
where $\xi_{\bm k,s} = v_F|k| - \mu$, is the electron
energy relative to 
the chemical potential $\mu$ ($= E_F$ at $T=0$).
The function $N_0$ and $n_F$ are Bose and Fermi distribution function
defined by
\begin{equation}
N_0 = \frac{1}{e^{\beta \omega_{SO}} -1},
\end{equation}
and
\begin{equation}
n_F(k_{s}) = \frac{1}{e^{\beta \xi_{ks}}+1},
\end{equation}
where the chemical potential $\mu$ at finite temperature is determined from the total
electron density $n$ of the system. 
The self energy is an explicit function of the independent variable
momentum ($k$) and energy $\omega$.
We obtain the real and imaginary parts of the electronic self
energy
\begin{widetext}
\begin{equation}
{\rm Re}[\Sigma_s(\bm k,\omega)] =   M_0^2\sum_{s'}\int\frac{k'dk'
 }{(2\pi)^2} G_{ss'}(\bk,\bk')  \left [ \frac{N_0 + 
      n_F(\xi_{k's'})}{\omega  + \omega_{SO} - \xi_{k's'}} + \frac{N_0 +1
      -n_F(\xi_{k's'})}{\omega  -\omega_{SO}-\xi_{k's'}} \right ],
\label{eq:re}
\end{equation}
\begin{equation}
{\rm Im}[\Sigma_s^(\bm k,\omega)]  =  -\pi M_0^2 \sum_{s'}\int\frac{k'dk'
 }{(2\pi)^2} G_{ss'}(\bk,\bk') [ (N_0 + 
      n_F(\xi_{k's'})) \delta(\omega + \omega_{SO} - \xi_{k's'}) 
 +  (N_0 +1
      -n_F(\xi_{k's'}))\delta(\omega -\omega_{SO}-\xi_{k's'})  ],
\label{eq:im}
\end{equation}
\end{widetext}
where
\begin{equation}
G_{ss'}(\bk,\bk') = \int d\theta \frac{e^{-2|\bk-\bk'|d}}{|\bk - \bk'|
  \epsilon(|\bk-\bk'|)} F_{ss'}(\bk,\bk'),
\label{eq:ggq}
\end{equation}
where $\theta$ is the angle between $\bk$ and $\bk'$.

For $T=0$ the imaginary part of the self energy,
Eq.~(\ref{eq:im}), becomes
\begin{widetext}
\begin{eqnarray}
{\rm Im}[\Sigma_{s}(k,\omega)] = A \sum_{s'=\pm1} \left \{
G_{ss'}[k,s'(\omega+\omega_{SO}+E_F)]  
s'(\omega+\omega_{SO}+E_F)\theta(-\omega - \omega_0)\theta[s'(\omega
    + \omega_{SO}+E_F)] \right . \nonumber \\
\left . +
G_{ss'}[k,s'(\omega-\omega_{SO}+E_F)]s'(\omega-\omega_{SO}+E_F)\theta(\omega
- \omega_0)\theta[s'(\omega 
    - \omega_{SO}+E_F)] \right \},
\label{eq:imt0}
\end{eqnarray}
\end{widetext}
where $A = -\pi M_0^2/(2\pi)^2 v_F$ and $\theta(x) =1$ for $x \ge 0$
and 0  for $ x < 0$. The first term in Eq.~(\ref{eq:imt0}) denotes
SO phonon absorption by the electron, whereas the second
term denotes the emission of SO phonon.
From Eq.~(\ref{eq:imt0}) we find that the phonon emission is only allowed
for $\omega > \omega_{SO}$, and the phonon absorption for $\omega < -\omega_{SO}$. 
Thus, due to the phase space restriction
there is a region where
Im$[\Sigma(k_F,\omega)]$ becomes zero for energies within
$\omega_{SO}$ of the Fermi energy, i.e.,
for $-\omega_{SO} < \omega < \omega_{SO}$ Im$[\Sigma(k_F,\omega)] =0$.
Note that $\omega=0$ corresponds to the Fermi energy.
The vanishing of Im$[\Sigma]$ indicates that a quasiparticle within
this region cannot decay by emitting (or absorbing) surface phonons because
there is no available final state to decay. 
A quasiparticle with an initial state energy $\omega$
has a final state energy $\omega - \omega_{SO}$ by emitting an SO
phonon. The final state energy of the electron must greater than the Fermi
energy (i.e., $\omega-\omega_{SO} > 0$) 
for a decay process to occur. Otherwise
the transition cannot occur because the final states are all occupied. 
Thus, for the quasiparticle with energy $\omega  < 
\omega_{SO}$, Im$[\Sigma]=0$. 
An electron outside this energy range can lose energy by emitting SO phonons.
For a quasiparticle with the initial energy $\omega$ has
the final state $\omega+\omega_{SO}$ by absorbing an SO phonon.  
Thus, the quasiparticle with the initial energy $-\omega_{SO} <
\omega$  cannot decay by absorbing an SO phonon.  
Thus, for $-\omega_{SO} < \omega < \omega_{SO}$, Im$[\Sigma(k_F,\omega)] =0$.
In addition to this
forbidden region, the decay of an electron to the Dirac point is
not allowed because the DOS is zero at that point, which occurs at
$\omega=-E_F-\omega_{SO}$ (i.e., the quasiparticle with the initial energy
$-E_F-\omega_{SO}$ cannot decay to the final state energy $-E_F$
(i.e. Dirac point) after absorbing an SO phonon).
However, we note that this forbidden gap exists only for the self-energy
contribution from surface optical 
phonons and at zero temperature. The quasiparticle will have finite values
of Im[$\Sigma$] in this energy region  
arising from scattering by impurities and from electron-electron 
interactions.

From the self energy, $\Sigma(\bk,\omega)$, we can obtain the
single-particle spectral function
\begin{equation}
A(k,\omega) = -2 {\rm Im} G(k,\omega)
\end{equation}
where
\begin{equation}
[G(k,\omega)]^{-1}=[G_0(k,\omega)]^{-1} - \Sigma(k,\omega)
\end{equation}
Thus, we have
\begin{equation}
A(k,\omega) = -\frac{ 2{\rm Im}\Sigma(k,\omega)}{\left [
    \omega-\xi_k-{\rm Re}\Sigma(k,\omega) \right ]^2 + \left [ {\rm
      Im} \Sigma(k,\omega) \right ]^2}.
\end{equation}
The spectral function $A(k,\omega)$ can roughly be interpreted as the
probability density of the different energy eigenstates required to
make up a specific $k$ state. The spectral function shows a sharp peak
at the quasiparticle state and becomes a $\delta$ function when
Im$[\Sigma] = 0$. The spectral function must satisfy the sum rule
\begin{equation}
\int \frac{d\omega}{2\pi}A(k,\omega) = 1,
\end{equation}
and  in all our numerical calculations 
we have explicitly checked that this sum rule
is generally satisfied to within less than a percent.

The inverse quasiparticle lifetime (or, equivalently, the scattering
rate) $\Gamma_s(\bm{k})$ of state $|s\bm k\rangle$  
is obtained by setting the frequency in imaginary part of the
self-energy to the on-shell (bare quasiparticle) energy $\xi_{s\bm
  k} = sv_F |{\bm k}| - E_F$, {\em i.e.}  
\begin{equation}
\Gamma_s(\bm{k}) = 2\,{{\rm Im}}[\Sigma_s^{\rm ret}(\bm{k},\xi_{\bm
    ks})].
\end{equation}
The self-energy approximation
used  here is equivalent to the generalized Born approximation for the scattering
rate.  Note that the integrand of Eq.~(\ref{eq:im}) is non-zero only when
$\xi_{ks} = \pm \omega_{SO} + \xi_{k's'}$, which
correspond to emission $(+)$ and absorption $(-)$ of surface optical
phonon, respectively.

\section{results}

In this section we present the results for graphene on SiC. The
parameters used in this calculation are following: the surface optical phonon frequency
$\omega_{SO}=117.6$ meV, the longitudinal optical phonon frequency,
$\omega_{LO}=120.7$ meV, the static (high frequency) dielectric
constant $\epsilon_0=9.72$ ($\epsilon_{\infty}=6.52$).

\begin{figure}[t]
\includegraphics[width=1.\columnwidth]{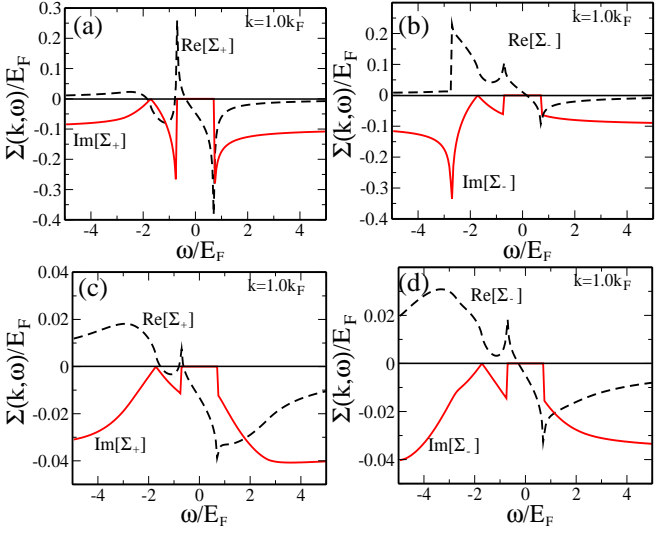}
\caption{Calculated self energies as a function of energy,
  $\omega/E_F$. In (a) and (b)
  $\Sigma_{\pm}(k,\omega)$ is calculated
  without screening effect [i.e. $\epsilon(q) = 1$ in
    Eq.~(\ref{eq:2})] and for
  $k=k_F$, $d=0$, and $T=0$. In (c) and (d) the self-energy 
  $\Sigma_{\pm}(k,\omega)$ is calculated
  with RPA screening function. Solid red (dashed black) lines indicate 
  imaginary (real) parts of the self energy.
}
\label{fig:1}
\end{figure}

In Fig.~\ref{fig:1} we show the calculated self-energy as a function
of energy ($\omega/E_F$) for $k=k_F$, $d=0$, and $T=0$. Note that
$\omega=0$ and $-E_F$ represent the Fermi energy and Dirac point,
respectively. The density $n=2\times 10^{12}$ cm$^{-2}$ is used in the
calculation, which corresponds to $E_F = 165$ meV.  
The Fermi momentum ($k_F$) and the Fermi energy ($E_F$, relative to
the Dirac point energy)  
of graphene are given by $k_F = (4\pi n/g_s g_v)^{1/2}$ and $|E_F| =
v_F k_F$ where 
$n$ is the 2D carrier (electron or hole) density.
$\Sigma_{+}$ ($\Sigma_{-}$) represents 
the self energy for $s=1$ ($s=-1$). 
In Fig.~\ref{fig:1} (a) and (b) the self-energies are calculated
without screening effect by putting $\epsilon(q) = 1$ in Eq.~(\ref{eq:2}), but
in (c) and (d) the RPA screening function is used to calculate the self-energies.
As discussed in Sec. II, Im[$\Sigma_s$] vanishes for $-\omega_{SO} < \omega <
\omega_{SO}$ and at $\omega = -E_F - \omega_{SO}$, and the range of
the forbidden gap is independent of screening. 
For $s=-1$ there is a sharp strong peak in Im[$\Sigma_-$].
This singular behavior for unscreened
self energy for $s=-1$ is unphysical and arises from the long range nature of
the electron- SO phonon interaction, i.e., $v_{ph} \propto
1/q$.  The RPA screening makes two significant
modifications to the self energy compared with the results with the
unscreened electron-SO phonon interaction. The screening effects reduce the 
magnitude of both Re[$\Sigma$] and Im[$\Sigma$] by roughly a factor of 10,
and removes the logarithmic singularity in Im$\Sigma_-$ occurring at $\omega =
-E_F -\omega_{SO} - v_Fk$. 
Since screening makes the long range electron-phonon
interaction finite, the singularity at $q=0$ (or, $k'=k$) is
removed by changing $q$ to $q\epsilon(q)$ in Eq.~(\ref{eq:imt0}).

\begin{figure}
\includegraphics[width=1.\columnwidth]{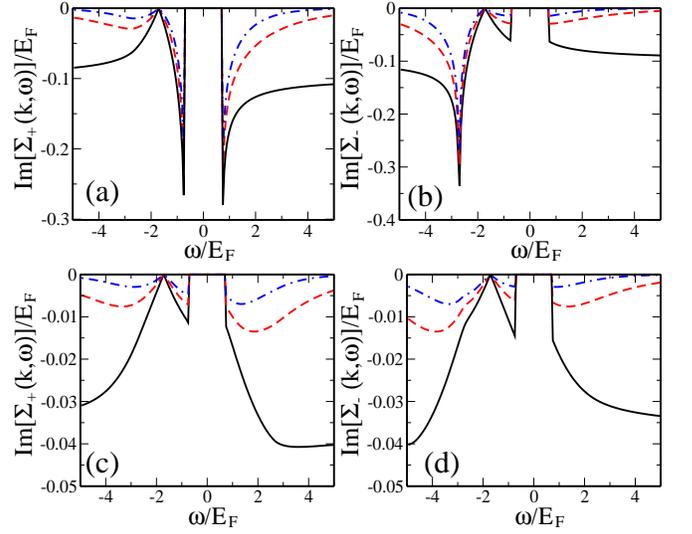}
\caption{Imaginary part of the self energy as a function of energy for
  $k=k_F$ and $T=0$ and for different values of distance between
  graphene and the dielectric substrate, $d = 0$ (solid line), 1nm
  (dashed line), 2nm (dot-dashed line) 
(a),(b) are calculated without screening effect and (c),(d)
  with RPA screening function.}
\label{fig:2}
\end{figure}

In Fig.~\ref{fig:2} Im$[\Sigma_{\pm}(k,\omega)]$ is shown 
as a function of energy $\omega$ for
$k=k_F$ and $T=0$, and for different values of $d=0$, 1, and 2 nm,
where $d$ is the distance between graphene and the interface of the
dielectric substrate. In Fig.~\ref{fig:2}(a)(b) the self-energy is 
calculated without screening effects 
and in (c) (d) the RPA screening in electron-phonon interaction is considered.
As the distance $d$ increases the magnitude of the Im[$\Sigma$] decreases due to
the exponential dependence of the interaction, i.e., $e^{-2qd}$ in
the electron-SO phonon interaction, [see 
Eqs.~(\ref{eq:mq}) and (\ref{v_phss})].
In Fig.~\ref{fig:2}(a) and (b) the logarithmically diverging peaks at
$\omega = \pm \omega_{SO}$ and $\omega = -2k_F + \omega_{SO}$
arises from the singular behavior of the long range electron-SO phonon
interaction at $q=0$. However, as discussed in Fig.~\ref{fig:1}, this
peaks disappear in the presence of screening as shown in
Fig.~\ref{fig:2} (c) and (d).
For comparison of these results with the imaginary parts of the
self-energy calculated with 
non-polar in-plane LO phonon ($\omega_{LO} \sim 200$ meV) arising from
the lattice vibration of 
graphene itself which has been
discussed in Ref.~[\onlinecite{tse2007}] and
[\onlinecite{calandra2007}], we consider the self-energy for 
$d=0$ and the unscreened case ($\epsilon(q) = 1$). 
Then, the main difference between these two cases appears in
Eq.~(\ref{eq:imt0}). For non-polar LO phonon we have
$G_{ss'} = \pi$, and for polar SO phonon we have 
\begin{eqnarray}
G_{s,s}(k,k') & = & \frac{k+k'}{kk'}\left [ K(r) - E(r) \right ]
\nonumber \\
G_{s,-s}(k,k') & = & \frac{k+k'}{kk'}\left [- \left (
  \frac{k-k'}{k+k'} \right )^2 K(r) + E(r) \right ],
\end{eqnarray}
where $K(r)$ and $E(r)$ are the complete elliptic integral of the
first and second kinds, respectively, and $r = 2kk'/(k^2 + k'^2)$.
For the non-polar LO phonon Im[$\Sigma$] increases linearly with
$|\omega|$ for larger $|\omega|$ because the DOS increases linearly with 
$|\omega|$. For polar SO phonon the long range interaction and the
linear energy dispersion of graphene DOS compensate each other and the
cancellation of these two effects makes the 
self energy saturate for large $|\omega|$. 
The emission (or absorption) of SO
phonon does not depend on the quasiparticle energy for large
$|\omega|$ (our numerical results show that the Im[$\Sigma$] saturates for
$|\omega| \agt 4 E_F$).  
Thus, high energy hot electrons
decay more effectively by emitting (absorbing) graphene LO phonons than
low energy electron. However, for SO phonons the damping of the
electron by emitting (absorbing) SO phonons depends weakly on the energy
of electrons. For finite $d$,  Im[$\Sigma$] decreases faster than for $d=0$ as
$|\omega|$ increases, then the high energy hot electrons do not decay effectively by
emitting (or absorbing) remote SO phonons.

\begin{figure}
\includegraphics[width=1.0\columnwidth]{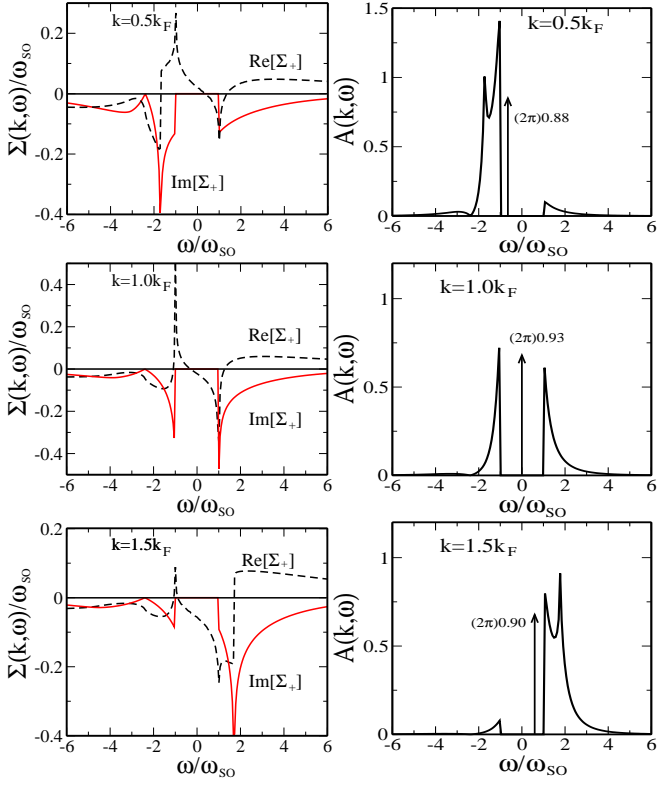}
\caption{Calculated self energy $\Sigma_+$ and spectral function $A_+$
  as a function of energy for 
  different wave vectors, $k=0.5$, 1.0, 1.5 $k_F$ and for $d=$ 1nm and
  $T=0$. Note that the 
  energies are measured in terms of $\omega_{SO}$ instead of $E_F$. 
Screening effect is
  not considered, $\epsilon(q) = 1$. The vertical arrows indicate
$\delta$-function peaks at the quasiparticle energy with the spectral
  weights shown next to the arrows. They are located
  at $\omega = -0.447$, 0.0, and 0.454 $E_F$ for $k=0.5$, 1.0, 1.5
  $k_F$, respectively. We find that from the calculated spectral function
the renormalized velocity due to the electron-SO phonon interaction 
is reduced, i.e., $v_F^*= 0.88 v_F$ approximately for these results. 
}
\label{fig:3}
\end{figure}
\begin{figure}
\includegraphics[width=1.0\columnwidth]{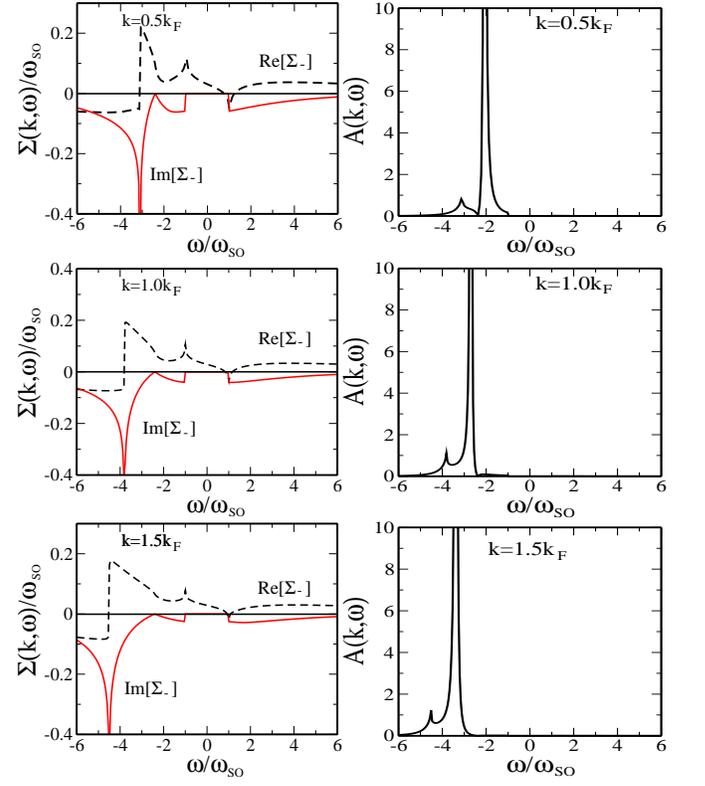}
\caption{Self energy $\Sigma$ and spectral function $A$ for $s=-1$
calculated with unscreened electron-SO phonon interaction.
$\Sigma_-$ and $A_-$ are shown 
as a function of energy $\omega/\omega_{SO}$ for
different wave vectors, $k=0.5$, 1.0, $1.5k_F$ and for $d=1$ nm and $T=0$.
The quasiparticle
  peaks are not coherent for $s=-1$ and the peaks are broadened by
  emitting or absorbing SO phonons.
}
\label{fig:33}
\end{figure}

\begin{figure}
\includegraphics[width=1.0\columnwidth]{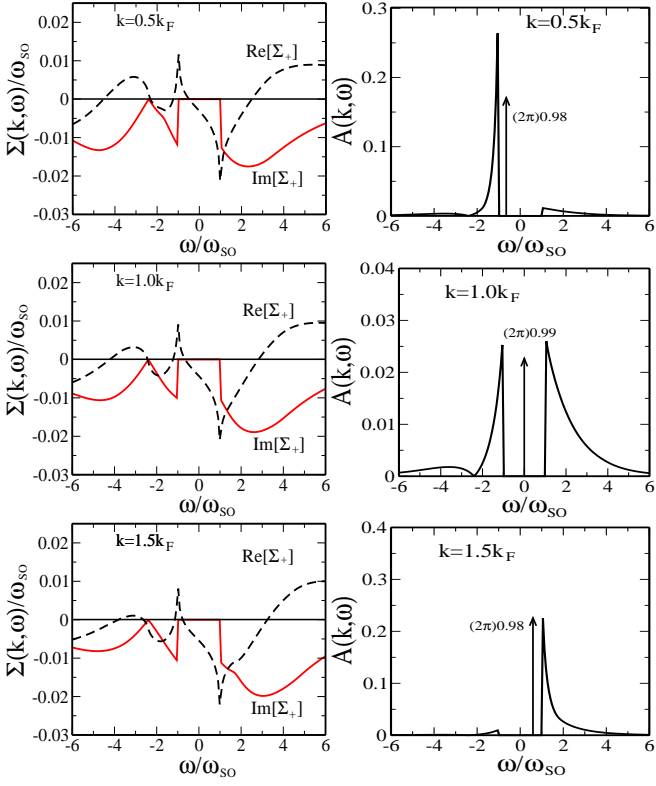}
\caption{Calculated self energy $\Sigma_+$ and spectral function $A_+$
  as a function of energy for 
  different wave vectors, $k=0.5$, 1.0, $1.5k_F$ and for $d=1$ nm and
  $T=0$. RPA screening  
  is considered in the electron-SO phonon interaction. Arrows indicate
$\delta$-function peak at the quasiparticle energy.
The renormalized velocity due to the screened electron-SO phonon interaction 
is bareley reduced, giving $v_F^*= 0.98 v_F$ approximately for these results. 
}
\label{fig:4}
\end{figure}

\begin{figure}
\includegraphics[width=1.0\columnwidth]{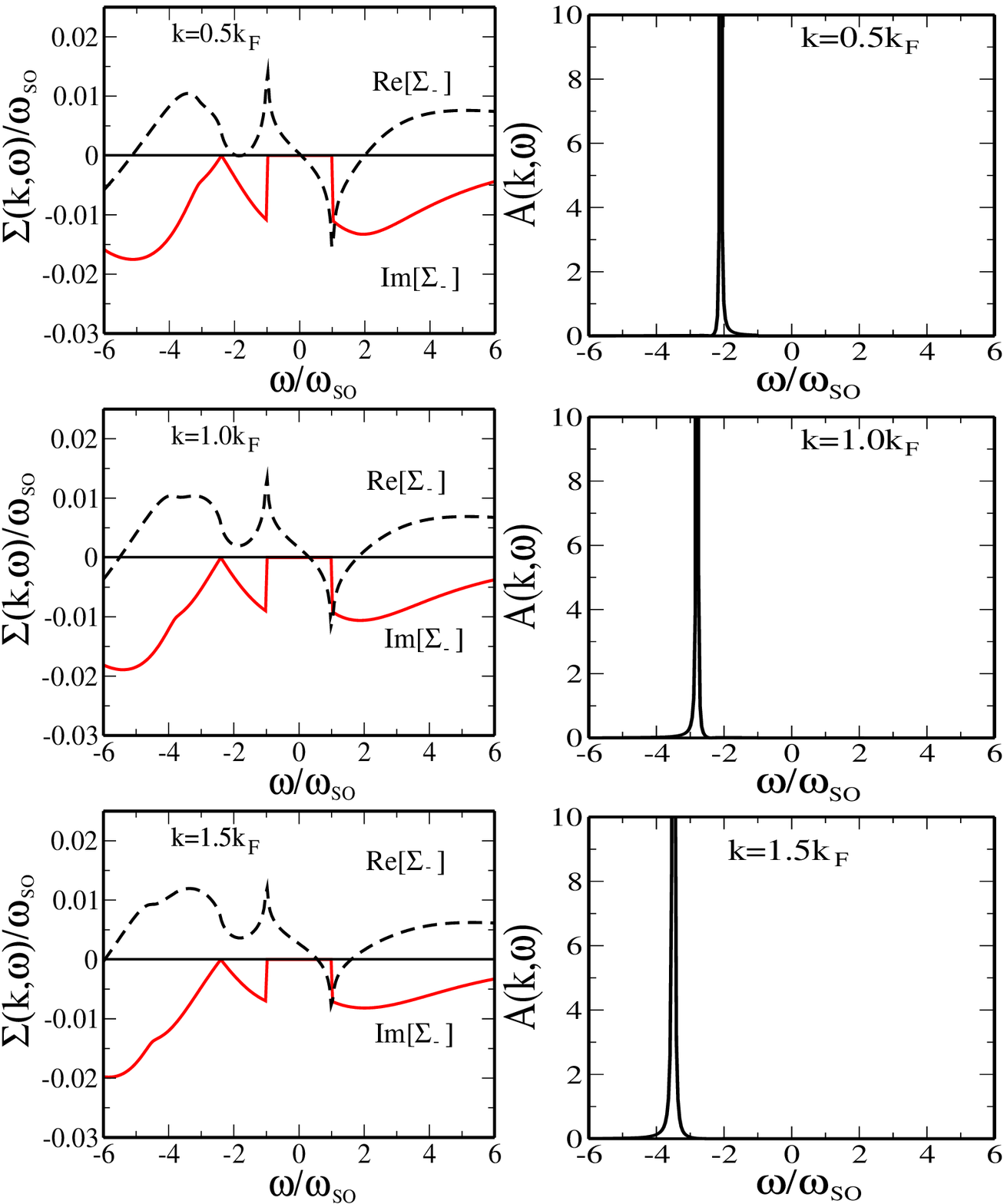}
\caption{Calculated self energy $\Sigma_-$ and spectral function $A_-$
  as a function of energy for 
  different wave vectors, $k=0.5$, 1.0, $1.5k_F$ and for $d=1$ nm and
  $T=0$. RPA screening effect 
  is considered in the electron-SO phonon interaction.
}
\label{fig:44}
\end{figure}

In Figs.~\ref{fig:3}---\ref{fig:44} we show the calculated self-energy
$\Sigma(k,\omega)$
and the spectral function $A(k,\omega)$ for 
different values of $k = 0.5$ 1.0, and 1.5 $k_F$. 
Figs.~\ref{fig:3} and \ref{fig:33} show the results calculated with unscreened
electron-SO phonon interaction for $s=+1$ (conduction band) and
$s=-1$ (valence band), respectively. The bare electron (hole) energy, $E(k) = \pm v_F
|k|$, is modified due to the electron-SO 
phonon interaction and the
energy dispersion is given by the solution to
Dyson's equation for a given $k$,
\begin{equation}
\omega = \xi_k + {\rm Re}[\Sigma_s(k,\omega)] -\mu,
\end{equation}
where $\xi_k = v_F |k| - E_F$ and $\mu = \Sigma_+(k_F,0)$ is the chemical potential.
A peak in the spectral function corresponds to the solution to
Dyson's equation and the quasiparticle energy for a given $k$.
For $s=+1$ (Fig.~\ref{fig:3}) the spectral function has a well
defined $\delta$-function peak as long as the peak is located inside
the forbidden region. 
The vertical arrows in Fig.~\ref{fig:3} indicate
$\delta$-function peaks at the quasiparticle energy with the spectral
  weights shown next to the arrows. They are located
  at $\omega = -0.447$, 0.0, and 0.454 $E_F$ for $k=0.5$, 1.0, 1.5
  $k_F$, respectively, and these 
quasiparticle peaks consume almost 90\% of the total spectral weight.
The remaining part of the spectral weight goes to the incoherent background.
We find that from the calculated spectral function
the renormalized velocity due to the electron-SO phonon interaction 
is reduced, i.e., $v_F^*= 0.88 v_F$ approximately for these results.
In Fig.~\ref{fig:33} we show the self-energy $\Sigma_-(k,\omega)$
and the spectral function $A_-(k,\omega)$ for $s=-1$ and for
different values of $k = 0.5$ 1.0, and 1.5 $k_F$. For $s=-1$ the
quasiparticle peaks are located outside the forbidden region and they
decay by emitting (or absorbing) SO phonons. As a consequence, the
spectral function has a peak 
with finite broadening instead of the $\delta$-function peak.
The quasiparticle peaks for $s=-1$ (or in the valence band) are not
coherent and the peaks are broadened by emitting or absorbing SO
phonons.

Figs.~\ref{fig:4} and \ref{fig:44} show the results calculated with
the screened
electron-SO phonon interaction for $s=+1$ (conduction band) and
$s=-1$ (valence band), respectively. 
For $s=+1$ we have the $\delta$ function peak
in the spectral function corresponding to the solution to 
Dyson's equation and the quasiparticle energy for a given $k$.
However, the energy of the quasiparticle (i.e., the location of the
peak) for a given wave vector is
almost identical to that of the non-interaction systems.
The peaks are located
  at $\omega = -0.49$, 0.0, and 0.49 $E_F$ for $k=0.5$, 1.0, 1.5
  $k_F$, respectively, and these 
quasiparticle peaks consume almost all ($99\%$) of the total spectral
weight. Just less than 1\% of the spectral weight goes to the
incoherent background. 
The renormalized velocity due to the screened electron-SO phonon interaction 
is barely reduced, giving $v_F^*= 0.98 v_F$ approximately for these results. 
Since screening reduces the
magnitude of the imaginary part of the self energy as shown in
Figs.~\ref{fig:1} and \ref{fig:2} the spectral weight of the
incoherent background is diminished in the presence of
screening, which gives rise to the increase of the
spectral weight in the quasiparticle peak because of the sum rule.
In Fig.~\ref{fig:44} we show the self-energy $\Sigma_-(k,\omega)$
and the spectral function $A-(k,\omega)$ calculated with the screened
electron-SO phonon interaction.
For $s=-1$ the quasiparticle peaks in the presence of screening are also
incoherent and the peaks are broadened by emitting or absorbing SO
phonons. However, the reduction of the magnitude of the
self-energy due to screening the broadened quasiparticle peaks become much sharper
than the peaks without screening.

\begin{figure}
\includegraphics[width=1.0\columnwidth]{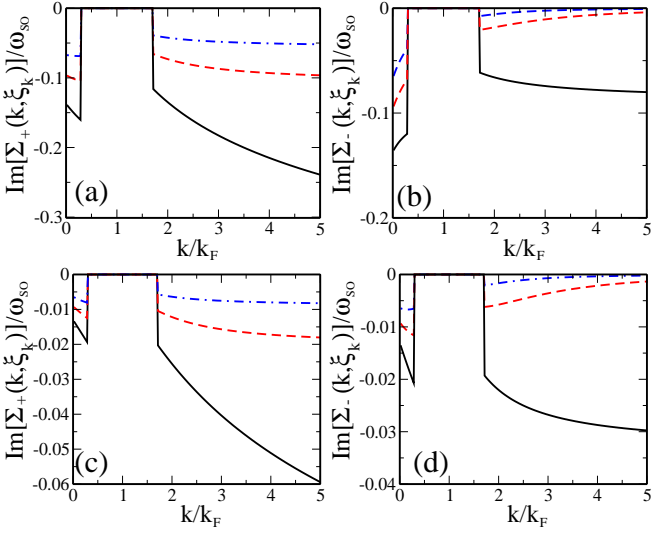}
\caption{The imaginary part of the on-shell self energy,
  Im[$\Sigma(k,\xi_k)$], as a function of wave vector for different
 $d=0$ (solid lines), 1 (dashed lines), 2 nm (dot dashed
  lines). The calculation is done at $T=0$.
(a) and (b) are the results with unscreened electron-phonon interaction for $s=+1$ and
  $s=-1$, respectively. (c) and (d) are results with screening
  interaction for $s=+1$ and $s=-1$, respectively.   
The forbidden gap, where
Im$[\Sigma_s(k,\xi_k) = 0]$, is given by $k_F - \omega_{SO}/v_F < k <
k_F + \omega_{SO}/v_F$. 
} 
\label{fig:7}
\end{figure}

In Fig.~\ref{fig:7} we show the imaginary part of the on-shell self energy, i.e.,
the self energy at $\omega = \xi_k= v_F k -E_F$. 
For $0< k < k_F-\omega_{SO}/v_F$ electrons decay by absorbing SO
phonons, while for $k > k_F + \omega_{SO}/v_F$ electrons decay by
emitting SO phonons. Thus, Im$[\Sigma_s(k,\xi_k)] = 0$ for $k_F -
\omega_{SO}/v_F < k < k_F + \omega_{SO}/v_F$. 
Inside the forbidden gap, the electron
does not decay by emitting or absorbing a SO phonon.
For $d=0$ and large wave vectors $k$ we have Im$[\Sigma_+(k,\xi_k)]
\propto G_{++}(k,k-\omega_{SO}) 
(k-\omega_{SO}) \propto \log(k)$ 
and Im$[\Sigma_-(k,\xi_k)] \propto G_{-+}(k,k-\omega_{SO})
(k-\omega_{SO}) \sim 2$. The Im$[\Sigma_+(k,\xi_k)$
  increases logarithmically and Im$[\Sigma_-(k,\xi_k)$
    saturates for large $k$. Thus, the damping of electrons by emitting SO
    phonon depends weakly on the energy of electrons. However, 
the Im[$\Sigma_+(k,\xi_k)$] for the non-polar LO phonon 
increases linearly with wave vector and high energy electrons decay
effectively by emitting more LO phonons \cite{tse2007,calandra2007}.
Screening effects reduce the magnitude of the on-shell energy as
shown in Fig.~\ref{fig:7}(c) and (d).
In addition to the reduction due to screening 
the magnitude of the imaginary part of the on-shell self-energy
decreases with increasing $d$ (the distance 
between graphene and the interface of the
dielectric substrate). 

\begin{figure}
\includegraphics[width=1.\columnwidth]{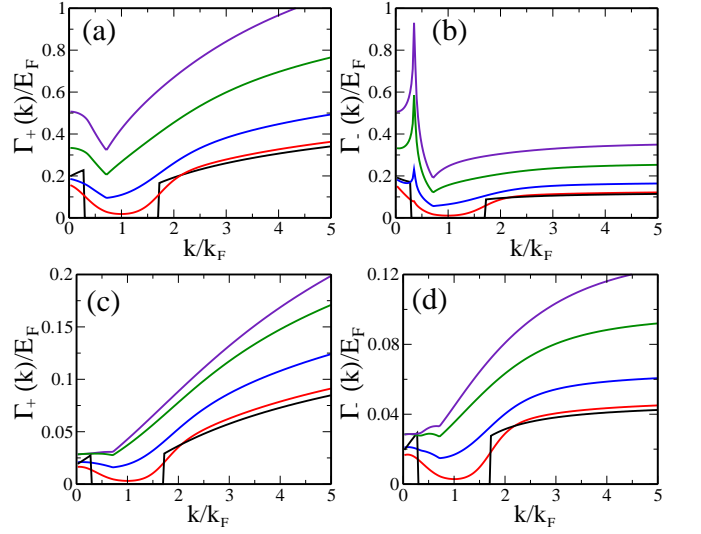}
\caption{Calculated damping rate $\Gamma(k) = 2 {\rm
    Im}[\Sigma(k,\xi_k)]$ as a function of wave vector for
  different temperatures, $T=0$, 0.2, 0.4, 0.7, 1.0 $T_F$ (from bottom
  to top), where $T_F
  = E_F/k_B$ is the Fermi temperature. The carrier density $n=2\times
  10^{12} cm^{-2}$ and $d=0$ are used in this calculation.
(a) and (b) are the results with unscreened electron-phonon interaction for $s=+1$ and
  $s=-1$, respectively. (c) and (d) are results with screened
  interaction for $s=+1$ and $s=-1$, respectively.   
}
\label{fig:8}
\end{figure}

In Fig.~\ref{fig:8} the calculated damping rate, $\Gamma(k)=2 {\rm
Im}[\Sigma(k,\xi_k)]$, is shown as a function of wave vector for 
different temperatures. Here, we assume that the temperature of 
electron and phonon is the same.
As temperature increases the forbidden gap
disappears due to the thermal smearing of the Fermi surface, which
loosens the phase space restriction.
Thus, SO phonon emission (absorption) is allowed for $k < k_F +
\omega_{SO}/v_F$  ($k > k_F - \omega_{SO}/v_F$).
The quantitative effect of screening (compared with the unscreened
result) is about a factor of five reduction of the scattering rate as
shown in Fig.~\ref{fig:8} (c) and (d). In addition to the reduction of
the rate, screening effects are particularly important in
$\Gamma_-(k)$. In Fig.~\ref{fig:8} (b) the peak inside the forbidden
region at $T=0$ develops as temperature increases. The peak occurs at
$k=\omega_{SO}/2v_F$ and arises from the singular nature of the long
range bare electron-phonon interaction. The divergence of the
interaction at $q=0$ disappears in the presence of screening effects
and the developed peak in scattering rate also disappears due to
screening effects. 

\begin{figure}
\includegraphics[width=.8\columnwidth]{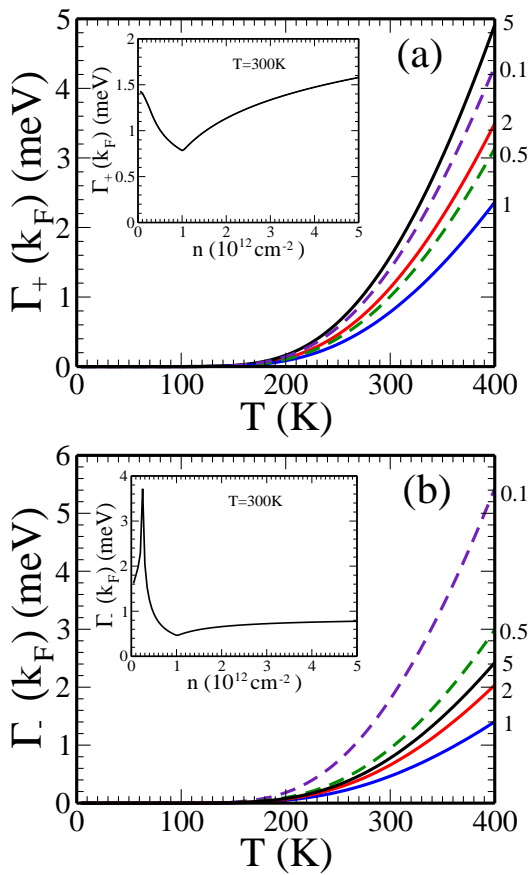}
\caption{Calculated damping rate at $k=k_F$ as a function of temperature for
  different electron densities, which are shown next to the lines in
  the unit of $10^{12}$ cm$^{-2}$. The results are calculated with
  $d=0$ and without screening effects.   
Insets show the calculated damping rate as a function of density at $T=300$K.
The density dependent damping rate have a local minimum at the density
corresponding to $E_F = \omega_{SO}$.
}
\label{fig:9}
\end{figure}

In Fig.~\ref{fig:9} the calculated damping rate at $k=k_F$,
$\Gamma(k_F)$, is shown as a function
of temperature for different electron densities.
The results in Fig.~\ref{fig:9} are calculated without screening
effects, but we notice that the screening effects reduce the
scattering rate approximately by a factor of five as shown in
Fig.~\ref{fig:9}.  
The scattering rate increases exponentially with temperature due to
the increases of phonon population, $\Gamma \propto e^{-\omega_{SO}/T}$
as temperature increases.
Insets in Fig.~\ref{fig:9} show the calculated damping rate as a
function of density at $T=300$K. 
The density dependent damping rate at a given temperature is
non-monotonic and has a local minimum at the density
corresponding to $E_F = \omega_{SO}$. Thus, the damping rate decreases
first as the density increases, and has a minimum at
$n=(\omega_{LO}/v_F)^2/\pi$. The damping rate increases as the density
increases further. For $s=-1$ the damping rate shows more complicated
density dependence. The local minimum occurs at the same density
for $s=+1$, but there is a sharp peak at low density.
The sharp peak in $\Gamma_-(k_F)$ arises from the unscreened
electron-phonon interaction, and disappears when we consider the
screening effects.

\begin{figure}
\includegraphics[width=.8\columnwidth]{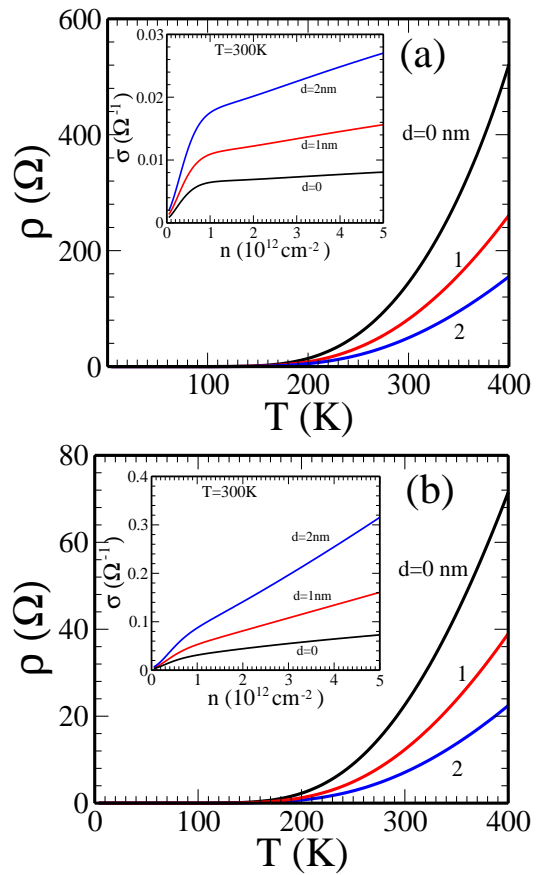}
\caption{Calculated resistivity $\rho = \sigma^{-1}$ as a function of
  temperature for
  different $d$, the distance between the dielectric interface and graphene.
The carrier density $n=2\times 10^{12}$ cm$^{-2}$ is used in this calculation.
In insets conductivity as a function of density for
  different $d$ is shown.
In (a) the screening effects is not considered, but in (b) the RPA
static screening is used.
The mobility is calculated by the simple formula
$\mu = 6.24 \times 10^5/\tilde{n} \rho$ cm$^2$/Vs where $\tilde{n}$ is the
carrier density ($\tilde{n} = n/10^{12}$cm$^{-2}$) and $\rho$
in unit of $\Omega$.
}
\label{fig:10}
\end{figure}

When the scattering rate $\Gamma$ is known, the conductivity can be
calculated by using Boltzmann transport equation \cite{dassarma2011,fratini2008} 
\begin{equation}
\sigma = \frac{e^2 v_F^2}{2} \int d \omega
\frac{D(\omega)}{\Gamma(\omega)} \left ( -\frac{\partial f}{\partial
  \omega} \right ),
\end{equation}
where $D(\omega) = g_sg_v \omega/(2\pi v_F^2)$ is the density of states of
graphene and $f$ is the Fermi distribution function.
In Fig.~\ref{fig:10} we show the calculated resistivity ($\rho =
\sigma^{-1}$)  as a function of temperature for different $d$, the
distance between the interface of the substrate and graphene. 
From the calculated resistivity the mobility is calculated by the simple formula
$\mu = 6.24 \times 10^5/\tilde{n} \rho$ cm$^2$/Vs where $\tilde{n}$ is the
carrier density measured in unit of $10^{12}$ cm$^{-2}$ and $\rho$
in unit of $\Omega$. The
calculated resistivity limited by the electron-SO phonon scattering
shows the exponential temperature dependence, which is the
characteristic behavior of the optical phonon scattering arising 
from the thermal occupation factor in the scattering rate.
The exponential temperature dependence is not affected by the
screening effects as shown in Fig.~\ref{fig:10}(b).
In insets of Fig.~\ref{fig:10} we show the calculated conductivity as a function
of density for different $d$. The density dependent conductivity
shows linear behavior, but the slopes are
different for low and high density regions. The 
characteristic electron density $n_c$ separating these two regions is given
by the density corresponding to the Fermi energy $E_F \sim
\omega_{SO}$. Below $n_c$ the conductivity increases fast but above
$n_c$ the conductivity increases slowly. The exponential temperature
dependence of the resistivity and  the linear density dependence of the
conductivity have recently been observed in the transport measurement
\cite{chen2008b,zou2010}.

\section{conclusion}

We theoretically study various aspects of the electron-surface optical
phonon interaction effects in graphene on
a substrate made of polar materials. 
We provide a rather complete set of numerical results of self-energy,
spectral function, and damping rate both for the unscreened interaction and
for the static RPA screening interaction. 

We find that the scattering rate with screened electron-SO phonon
interaction is reduced 
by a factor of five and the self-energy by a factor of ten, compared
with the results with unscreened  
interaction. The static RPA screening used in this calculation may overestimate the
screening  effect on scattering rate, but we believe that the static RPA
screening  provides an upper limit of the screening effects. 
That is, our unscreened and statically screened results for the surface polar
optical phonon
scattering rate provide two extreme bounds for the magnitude of the
scattering rate since static screening is an overestimation of the
actual dynamical screening.
We mention that the static screening approximation becomes very
accurate at high carrier density when the typical Fermi energy exceeds
the SO phonon energy, which may already happen at a rather low carrier
density of $10^{12}$cm$^{-2}$.  We therefore believe that our statically
screened results should be valid for most graphene samples used
experimentally.
In addition to the reduction of the magnitude of the self-energy the
screening effects remove the unphysical singular behavior in the self-energy
for $s=-1$ arising from the long range nature of the electron-SO
phonon interaction.

We find that the calculated renormalized Fermi velocity due to the
unscreened electron-SO phonon interaction is reduced (with respect to
the bare graphene velocity) by about 20\%, $v_F^* \sim 0.8 v_F$, but for the
screened interaction it is reduced by only 2\%,  $v_F^* \sim 0.98 v_F$ at a
density $n=2\times 10^{12}$ cm$^{-2}$. We note that the renormalized Fermi velocity
is enhanced about 20\% at the same density when the electron-electron
Coulomb interaction is only considered\cite{dassarma2013b}. Since our consideration
of screening in this paper is the RPA static dielectric function we
cannot state conclusively that the small change of the renormalized
velocity in the presence of screened electron-phonon interaction
arises from the direct cancellation between these two
interactions. More careful calculation is need to confirm this
cancellation by considering the dynamical screening effects treating
both the electron-SO phonon and the electron-electron Coulomb
interaction on an equal footing.  But it appears the graphene velocity
renormalization due to the electron SO phonon interaction by itself is
a rather small quantitative effect because of screening effect.  This
is certainly true at high carrier densities ($>10^{12}$ cm$^{-2}$) where
our static screening approximation should apply well. 
We also find that 
the scattering rate for the electron-SO phonon interaction depends weakly
on the electron energy due to the cancellation between the
long range nature of the interaction and the linear behavior of the
density of states of graphene. For the in-plane non-polar LO phonon of
graphene the scattering rate strongly depends on the electron energy
and high energy electrons are heavily damped by emitting LO
phonons. Since the scattering rate due to the SO phonon decreases
further due to the physical separation between graphene and interface
of the dielectric substrate the chance of the hot electron damping by emitting
SO phonons may be very low in the real graphene samples,
but the high-temperature transport in clean graphene may very well be
dominated by SO phonon scattering from the substrate.

\section*{acknowledgments}

This work was supported by U.S.-ONR.


\end{document}